\documentclass[onecolumn]{emulateapj}
\usepackage{amsmath}

\lefthead{Wada}
\righthead{Radiation-driven Fountain in AGNs}

\newcommand{\bea}{\begin{eqnarray} }
\newcommand{\eea}{\end{eqnarray}}
\def\mbf#1{\mbox{\boldmath ${#1}$}}

\begin{document}

\title{Radiation-driven Fountain and Origin of Torus around Active Galactic Nuclei}

\author{Keiichi \textsc{Wada}}%
\affil{Kagoshima University, Kagoshima 890-0065, Japan}
\email{wada@astrophysics.jp}

%




\begin{abstract}
We propose a plausible mechanism to explain the formation of the 
so-called ``obscuring tori'' around 
active galactic nuclei (AGNs) based on three-dimensional
hydrodynamic simulations including radiative feedback from the
central source.  
The X-ray heating and radiation pressure on the gas are explicitly 
calculated using a ray-tracing method.
This radiation feedback drives a ``fountain'',
that is, a vertical circulation of gas in the central a few to tens parsecs. 
Interaction between the non-steady outflows and inflows causes
the formation of a geometrically thick torus with internal turbulent motion. 
As a result, the AGN is obscured for a wide range of solid angles.
In a quasi-steady state, the opening angles for the column density toward a black hole $< 10^{23}$ cm$^{-2}$
are approximately $\pm 30^{\circ}$ and $\pm 50^{\circ}$ 
for AGNs with 10\% and 1\% Eddington luminosity, respectively.
Mass inflows through the torus coexist with the outflow and 
internal turbulent motion, and the average mass accretion rate to the central parsec region
is $2\times 10^{-4}\sim 10^{-3}\, M_{\odot}\; {\rm yr}^{-1}$; this is about
ten times smaller than accretion rate required to maintain
 the AGN luminosity.
This implies that relatively luminous AGN activity is intrinsically intermittent or that there are other mechanisms, such as stellar energy feedback, that enhance the mass accretion 
to the center.

\end{abstract}


\section{INTRODUCTION}

The presence of optically thick obscuring tori has been postulated to
explain the various observed properties of active galactic nuclei (AGNs), particularly with respect to
the two major categories of AGNs, namely,
type 1 and type 2 \citep{antonucci93}.  However, the origin of the 
optically and geometrically thick material that covers a large solid angle with respect to the AGN and the nature of its three-dimensional 
structure are still unclear,
although its geometry is often schematically described as a ``doughnut-shaped'' \citep[e.g.,][]{urry95}. 
In fact, statistical studies of the absorption features of local AGNs suggested have suggested the presence of complex,
inhomogeneous obscuring material around the central engine \citep{shi2006, hicks09}.
The clumpy torus model is also consistent with
the observed spectral energy distributions (SEDs) \citep[e.g.,][]{nenkova2002,elitzur06,levenson2009,schartmann09, stalevski2012}.
The essential question that then arises is regarding how such an inhomogeneous `torus' is formed and how it attains such a
large covering fraction
against the energy dissipation of the clumpy medium in the gravitational
field of the central black hole.

Numerous theoretical models have been proposed to explain the origin of the obscuring material.
On an accretion disk scale the proposed causes include
hydromagnetic disk winds \citep{konigl1994},
disk-driven hydromagnetic winds and radiation pressure due to continuum 
emission from the accretion disk \citep{keating2012}, and
warping of irradiated accretion disks \citep{arm97}.
Although a pasec-scale dusty torus is observed in some nearby AGNs \citep{jaffe04}, 
the nucleus could be obscured by relatively less dense ISM in the 
range of  10-100 pc or even by galactic disks themselves \citep{levenson2001,goulding2012}.
There are many instances where evidence suggests that most AGNs are associated with
circumnuclear starbursts \citep[e.g.][]{imanishi04, davies07, hicks09,chen09, imanishi2011, Diamond2012, Woo2012}. 
It is suggested that the star formation activities
drive mass inflow into AGNs, thereby leading to the growth of black holes  \citep[see the review by][]{alexander2011}.
The nuclear starbursts themselves could inflate the circumnuclear disk and 
obscure the central source. 
\citet{wada02} and \citet{wada09} have proposed that
a clumpy torus-like structure is naturally reproduced 
due to energy feedback from supernova explosions (SNe) in a disk.
In their model, the internal density and temperature of the thick disk are highly
inhomogeneous, and the velocity field of the disk is turbulent. 
The scale height of the disk is determined by the balance between the turbulent energy 
dissipation and energy input from SNe. 
More recently, \citet{hopkins2012} have proposed that 10-pc-scale thick tori 
can be formed even without the existence of strong stellar energy feedback if
gas inflow from kiloparsec scales drives instability in the circumnuclear region.

On a parsec scale or smaller, 
geometrically and
optically thick torus sustained by radiation pressure 
have been suggested \citep{kro88,pie92}.
Following this model, static solutions of thick disks supported by 
infrared radiation have been explored by \citet{krolik07} and \citet{shi-krolik2008}.
\citet{ohs01} have suggested that static, geometrically thin obscuring walls
could be formed by the interplay between the radiation from the nucleus and  
a ring-like nuclear starburst.


It would be more natural to assume that radiative feedback causes
the dynamical evolution of the surrounding ISM rather than static structures. 
\citet{roth2012} calculated 
the net rate of momentum deposition due to the radiation pressure
in the surrounding gas and estimated the
mass-loss rate by outflow using Monte Carlo radiative transfer calculations.
\citet{dorodnitsyn2012} performed `2.5D' (i.e. basic equations are solved in three dimensions with axisymmetry) radiative hydrodynamic simulations,
and they found that an AGN torus can be better described in 
terms of a radiationally supported flow rather than a quasi-static obscuring torus.
Using 2D simulations, \citet{schartmann2011} studied the evolution of dusty clouds irradiated by 
the AGN and found that the radiative transfer effect is significant depending on the column density of the clouds.

As is usual in multi-dimensional radiation-hydrodynamic simulations,  
cirtain simplifications have been applied in these previous studies, such as assuming flux-limited diffusion approximation and
introducing symmetry for the rotational axis and equatorial plane of the AGN.
In this study, we examine the effect of the radiation from the central source on the ISM 
by assuming a different set of simplifications.
We performed fully three-dimensional hydrodynamic simulations on scale of
a few tens pc around an AGN without introducing any symmetries.
We took into account radiative heating and pressure due to ``direct radiation'' from the central source along
rays for all the $256^{3}$ grid cells; however, we ignored the
phenomenon of scattering/re-emission.
This is a natural extension of our previous three-dimensional hydrodynamic simulations of tens-of-parsec scale tori \citep{wada02,wada05, 
yamada07,wada09},
in which self-gravity of the ISM, radiative cooling, SNe, and UV heating were taken into account.
Here, we report the first results without SNe feedback. The SED analysis based on the present model will appear elsewhere
(Schartmann and Wada, in prep.). 
In this study, we propose a
dynamical process leading to the formation of thick tori, and we
examine their structures in a quasi-steady state.
Further, we discuss the changes in the column density and the net mass accretion rate depending on the luminosity of the AGN.

\section{NUMERICAL METHODS AND MODELS}
\subsection{Numerical Methods}

Based on our three-dimensional, multi-phase hydrodynamic model \citep{wada09}, 
we included the radiative feedback
from the central source, i.e., radiation pressure on the dusty gas 
and the X-ray heating of cold, warm, 
and hot ionized gas. 
 In contrast to previous analytical and numerical studies of radiation-dominated tori\citep[e.g.,][]{pie92, krolik07,shi-krolik2008,Diamond2012,dorodnitsyn2011,dorodnitsyn2012}, we assume neither dynamical equilibrium nor geometrical symmetry.
Here, we solve fully three-dimensional hydrodynamic equations,
accounting for radiative feedback processes from the central source using a ray-tracing method.
In a manner similar to the approach by \citet{wada02}, we account for 
self-gravity of the gas, radiative cooling for 20 K $< T_{gas} < 10^{8}$ K, uniform UV radiation 
for photoelectric heating and H$_{2}$ formation/destruction \citep{wada09}; however, we do not include the effect of supernova feedback 
in this study,
in order to clarify the effect of the radiation feedback only. 
We plan to investigate
the effects of supernova in the circumnuclear starburst in a future study.

We use a numerical code based on Eulerian hydrodynamics with a uniform
grid \citep{wada_norman01,wada01,wada09}. 
\setcounter{footnote}{0}
We solve the following equations numerically to 
simulate the three-dimensional evolution of a rotating gas disk in a
fixed spherical gravitational potential under the influence of radiation from 
the central source.
  \begin{eqnarray}
{\partial \rho}/{\partial t} + \nabla \cdot (\rho \mbf{v}) &=& 0,
\label{eqn: rho} \\ 
{\partial (\rho \mbf{v})}/{\partial t} + (\mbf{v}
\cdot \nabla)\mbf{v}+{\nabla p}  &=&  - \rho(\nabla \Phi +\mbf{f}_{rad}^{r}), \label{eqn: rhov}\\
 {\partial (\rho E)}/{\partial t} + \nabla \cdot 
[(\rho E+p)\mbf{v}] -\rho \mbf{v}\cdot \nabla \Phi&=& 
\rho \Gamma_{\rm UV}(G_0) + \rho \Gamma_{\rm X}  -  \rho^{2} \Lambda(T_{gas}, f_{\rm H_2}, G_0)
, \label{eqn: en}\\ \nabla^2
\Phi_{\rm sg}&=& 4 \pi G \rho, \label{eqn: poi} 
\end{eqnarray}
 where $\Phi(\mbf{x}) \equiv \Phi_{\rm ext}(r) + \Phi_{\rm BH}(r)+\Phi_{\rm sg}(\mbf{x}) $;
$\rho,p,\mbf{v}$ denote the density, pressure, and velocity of
the gas, and
the specific total energy $E \equiv |\mbf{v}|^2/2+ p/(\gamma -1)\rho$,
with $\gamma= 5/3$. 
 We assume a time-independent external potential
$\Phi_{\rm ext}(r) \equiv -(27/4)^{1/2}[v_1^2/(r^2+
a_1^2)^{1/2}+v_2^2/(r^2+ a_2^2)^{1/2}]$, where $a_1 = 100$ pc, $a_2 =
2.5$ kpc, $v_1 = 147$ km s$^{-1}$, $v_2 = 147$ km s$^{-1}$, and
$\Phi_{\rm BH}(r) \equiv -GM_{\rm BH}/(r^2 + b^2)^{1/2}$, where $M_{\rm
BH}=1.3\times 10^7 M_\odot$ (see Fig. 1 in \citet{wada09} for 
the rotation curve).
The potential
caused by the black hole (BH) is smoothed within $r \sim b = 4\delta $, where
$\delta$ denotes the minimum grid size ($=0.25$ pc), in order to avoid too small time steps around
the BH. 
 In the central grid cells at $r
< 2\delta$, physical quantities remains constant.
  

We solve the hydrodynamic part of the basic equations using AUSM
 (Advection Upstream Splitting Method)\citep{liou03}.
We use $256^3$ grid points.
The uniform Cartesian grid covers a $64^3$ pc$^3$ region
around the galactic center (i.e. the spatial resolution is 0.25 pc).
The Poisson equation, eq.(\ref{eqn: poi}) is solved to calculate the
self-gravity of the gas using the fast Fourier transform and the
convolution method with $512^{3}$ grid points along wiht a periodic
Green's function.
We solve the non-equilibrium chemistry of hydrogen molecules
{along} with the hydrodynamic equations \citep{wada09}. 

We consider the radial component of the radiation pressure:
  \begin{eqnarray}
  \mbf{f}_{rad}^{r} = \int \frac{\chi_{\nu}\mbf{F}^{r}_{\nu}}{c} d\nu,
  \end{eqnarray}
where $\chi_{\nu}$ denotes the total mass extinction coefficient due to 
dust absorption and Thomson scattering, i.e. $\chi_{\nu} \equiv \chi_{dust,\nu}+\chi_{T}$.
The radial component of the flux at the radius $r$, $\mbf{F}^{r}_{\nu}$ is
  \begin{eqnarray}
  \mbf{F}^{r}_{\nu} \equiv \frac{L_{\nu}(\theta) e^{-\tau_{\nu}}}{4\pi r^{2}}\mbf{e}_{r},
  \end{eqnarray}
where $\tau_{\nu} =\int \chi_{\nu} \rho ds$. 

{The only explicit radiation source used here is an accretion disk 
whose size is 
five orders of magnitude smaller than the grid size in the present calculations.
Therefore we assumed that radiation is emitted from a point source.
However, the radiation flux originating from the accretion disk 
is not necessarily spherically symmetric \citep[e.g.][]{netzer1987}. 
In our study, we simply assume emission from a thin accretion disk without considering the limb darkening effect,
i.e., $L_{\nu} \equiv 2 L_{AGN}|\cos \theta|$, where $\theta$ denotes the angle from the rotational axis ($z$-axis)
(see also discussion in \S 4). 

In the simulations, we considered the radial component of the flux $\mbf{f}_{rad}^{r}$ only 
for the radiative heating and pressure.
This approximation is not necessarily correct especially with respect to optically thick dusty gas around 
the central few parsecs; however, beyond this region, the acceleration of the gas due to strong X-rays can be expected to
be radial. Even for the central region spanning a few parsecs, \citet{roth2012} showed that the gas acceleration is nearly radial using
three-dimensional Monte Carlo 
radiative transfer calculations. 
It is noteworthy
that since the hydrodynamics is calculated in a fully three-dimensional manner, the radiative acceleration
may contribute to gas dynamics along any direction. Therefore the resultant density and velocity structures 
could be very different from the symmetric ones \citep[cf.][]{ohs01}. }
The optical depth $\tau_{\nu}$ is calculated at every time step along a ray from the central source at
each grid point, i.e. $256^{3}$ rays are used in the computational box.
 
In the energy equation (eq. (3)), we use a cooling function based on a
radiative transfer model of photodissociation regions
\citep{meijerink05}, denoted as $\Lambda(T_g, f_{\rm H_2}, G_0)$, 
assuming solar metallicity, which is a function
of the molecular gas fraction $f_{\rm H2}$ and the intensity of far ultraviolet (FUV), $G_0$, \citep{wada09}.
 The radiative cooling rate below
$\simeq 10^4~{\rm K}$ is self-consistently {modified} depending on
$f_{\rm H_2}$, which is determined for each grid cell. We include
the effect of heating
due to { the} photoelectric effect, $\Gamma_{\rm UV}$.
We assume a
uniform FUV radiation field, with the Habing unit $G_0 =1000$,  originating in 
the star-forming regions in the central several tens parsecs.
%
In reality, both the density field and distribution of massive stars should 
not be uniform, {and} $G_0$ {should} have a large dispersion with some
radial dependence.  However, 
the FUV radiation affects the chemistry of the ISM (mostly distribution of molecular clouds) although this does not significantly alter
the dynamics of the circumnuclear ISM \citep{wada09}.
The effects of the non-uniform radiation field with multiple sources on the gas dynamics 
is an important problem, but it is beyond the scope of this paper.

The temperature of the interstellar dust, $T_{\rm dust}$,  at a given position irradiated by the central source
is calculated by assuming thermal equilibrium \citep[e.g.][]{nenkova2008}, 
and if $T_{\rm dust} > 1500$ K, we assume that the dust is sublimate.
Here we assume that the ISM in the central tens parsecs is optically thin 
with respect to the re-emission from the host dust.
The SED of the AGN and the dust absorption cross section are taken from \citet{laor1993}.
We use a standard galactic dust model, which is the same as that used by\citet{schartmann2010}, 
with eight logarithmic frequency bins between 
0.001 $\mu m$ and 100 $\mu m$.

The effects hard X-ray on the thermal and chemical structures of the ISM
are significant in the region
spanning subparsec distances to tens of parsecs \citep{jp2011}.
Further, X-ray heating may be essential for galactic scale feedback \citep{hambrick2011}.
Here, we consider heating by the X-ray, i.e.
interactions between the high energy secondary electrons
and thermal electrons (Coulomb heating), and H$_{2}$ ionizing heating
in the warm and cold gas \citep{maloney96, meijerink05}.
The Coulomb heating rate $\Gamma_{\rm X,c}$ for the gas
 with the number density $n$ is given by
\begin{eqnarray}
 \Gamma_{\rm X,c} = \eta n H_{\rm X},
\end{eqnarray}
where $\eta$ denotes the heating efficiency \citep{dalgarno1999,meijerink05}, and
 the X-ray energy deposition rate $H_{\rm X} = 3.8\times 10^{-25}\xi$ ergs s$^{-1}$ and
the ionizing parameter $\xi = 4\pi F_{\rm X}/n  = L_{\rm X}e^{-\int \tau_{\nu} d\nu}/n r^{2}$, 
with $L_{\rm X} = 0.1 L_{AGN}$.
For optically thin hot gas with $T \gtrsim 10^{4}K$, the effects of Compton heating and X-ray photoionization heating are
 considered, and the net heating rate \citep{Blondin1994}
 is approximately
\begin{eqnarray}
\Gamma_{\rm X,h}= 8.9\times 10^{-36} \xi (T_{\rm X} - 4T) +1.5\times 10^{-21} \xi^{1/4} T^{-1/2}(1-T/T_{\rm X}) \;\; {\rm erg}\, {\rm s}^{-1}\, {\rm cm}^{3},
\end{eqnarray}
with the characteristic temperature of the X-ray radiation, $T_{X} = 10^{8}$ K.

\subsection{Initial conditions and model parameters}
In order to prepare a quasi-steady initial condition without
radiation feedback, we first run a model of an axisymmetric and rotationally supported thin
disk with a uniform density profile (thickness of 2.5 pc) and a total
gas mass of $M_g = 6.68\times 10^6 M_\odot$.  Random density fluctuations, which are 
less than 1\% of the unperturbed values, 
are added to the initial gas disk.
%
%
After $t \sim 1.7$ Myr,  the thin disk settles in a flared shape
with the total mass $6.64 \times 10^6 M_\odot$. Since we allow outflows from
the computational boundaries, the total gas mass decreases during the
evolution.

A free parameter in the present simulations is the luminosity of the AGN.
Here, we explore models with two different Eddington ratios $L_{AGN}/L_{E} = 0.1$ and $0.01$,
where the Eddington luminosity $L_{E} \equiv  4\pi G c m_{p} M_{\rm BH}/\sigma_{T} $.
In each model,  $L_{\rm AGN}$ stays constant during calculations, i.e. $\sim$ 3 Myrs\footnote{
The numerical calculations were carried out on the NEC SX-9 (16 vector processors/1 TB memory/1.6 TFlops) at the 
National Astronomical Observatory of Japan and Cyberscience Center, Tohoku University. 
A single typical run for 3 Myr takes approximately 120 CPU hours.}.
%
\section{RESULTS}
%
\subsection{Onset of Thick Torus}
\begin{figure}[h]
\centering
\includegraphics[width = 11cm]{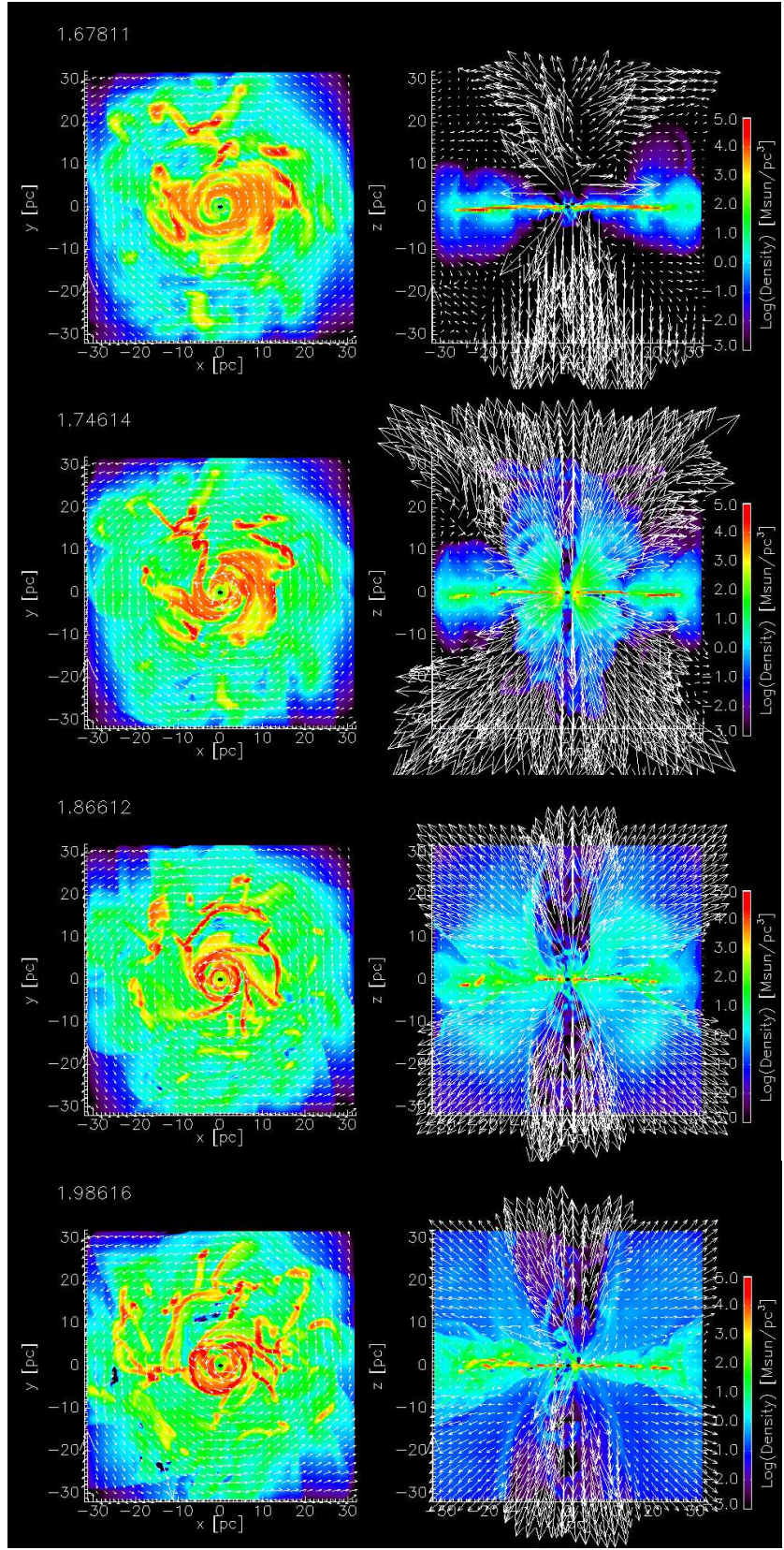} 
\caption{Initial evolution of radiation-driven outflow and onset of thick disk 
in a model with $L_{\rm AGN}/L_{\rm Edd} = 0.1$. {The vertical slices represent the $x$-$z$ planes.}
Arrows represent 
gas velocity (a unit vector of 500 km s$^{-1}$ is shown as reference along the $y$-axis.)}
\label{wada_fig: f1}
\end{figure}

Figure \ref{wada_fig: f1} shows the onset of biconical outflows, cavities, and
thick disk after the radiation from the central source is initiated.
Radiation-driven outflows start forming in the inner region, and subsequently propagate outward. At $t=1.986$ Myr, ``back-flows'' toward the disk plane appear
outside the biconical outflows. These accretion flows interact with the disk, 
thereby leading to the formation of a geometrically thick disk
that has a clear boundary with
 $\rho \sim 10^{2} M_{\odot}\; {\rm pc}^{-3}$ (Fig. \ref{wada_fig: f1}).

\begin{figure}[t]
\centering
\includegraphics[width = 14cm]{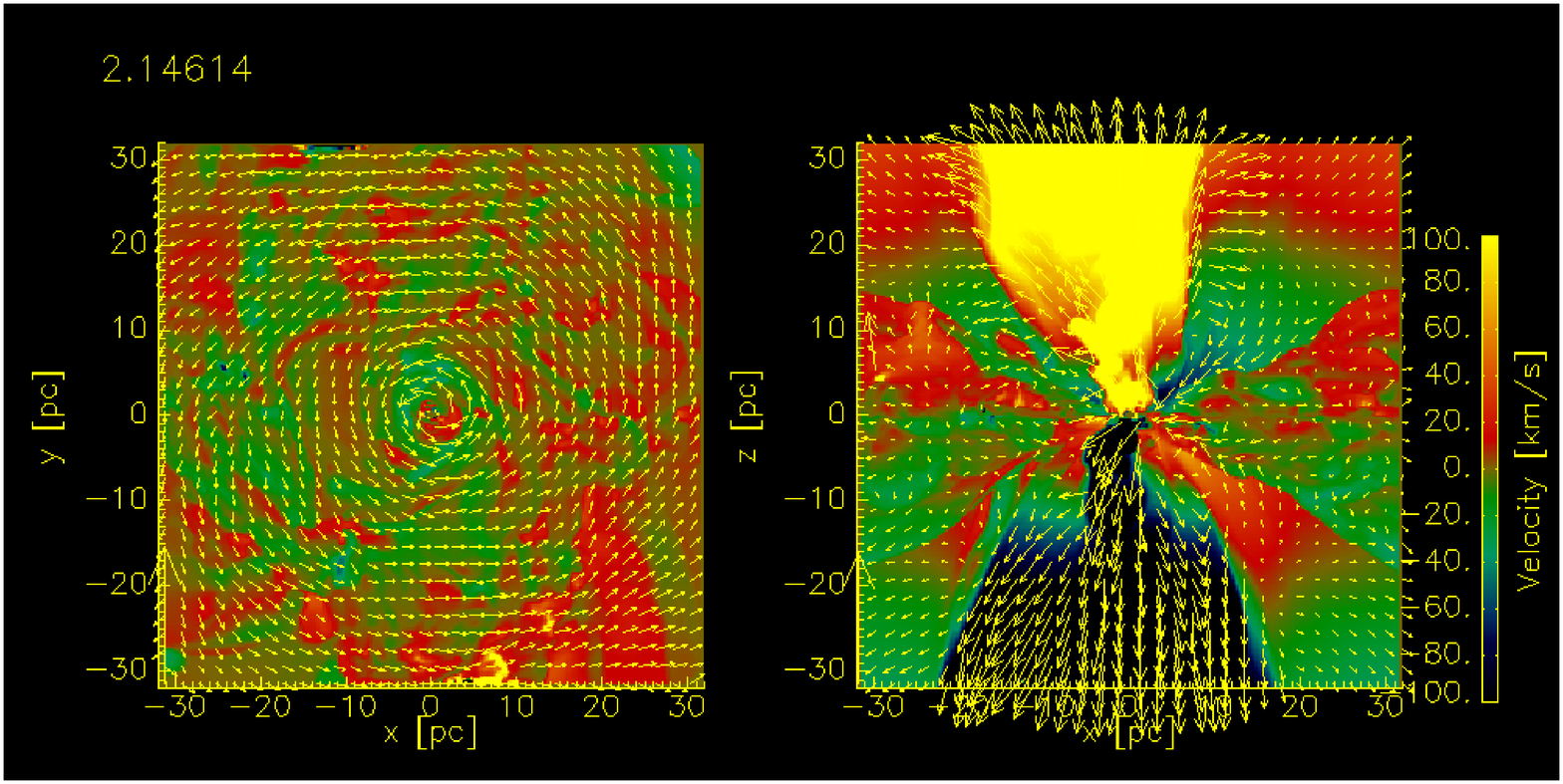} 
\caption{Observed $z$-component of velocity at $t=2.15$ Myr ($L_{AGN}/L_{E} = 0.1$). {The vertical slice indicates the $x$-$z$ plane.}
The downward flows (i.e., flow heading to 
the equatorial plane) appear outside the region of the biconical outflow, and these flows interact with 
the opposite directed flows, which are represented by the regions indicated in green/blue and red/yellow.
The upward and downward flows coexist and generate turbulent motions 
in the equatorial plane (left panel) as well as in the vertical direction.}
\label{wada_fig: f2}
\end{figure}

The interactions of vertical flows with opposite directions are more clearly seen in Fig. \ref{wada_fig: f2} ($t=2.146$ Myr), which 
shows the distribution of $v_{z}$.
There are several contact surfaces
between upward and downward flows. As a result, turbulent motions that
support the thick geometrical structure of the disk are generated. The complicated distribution of 
flows in the thick disk
is also seen in the velocity field at the equatorial plane (left panel of Fig. \ref{wada_fig: f2}).

\begin{figure}[t]
\centering
\includegraphics[width = 8cm]{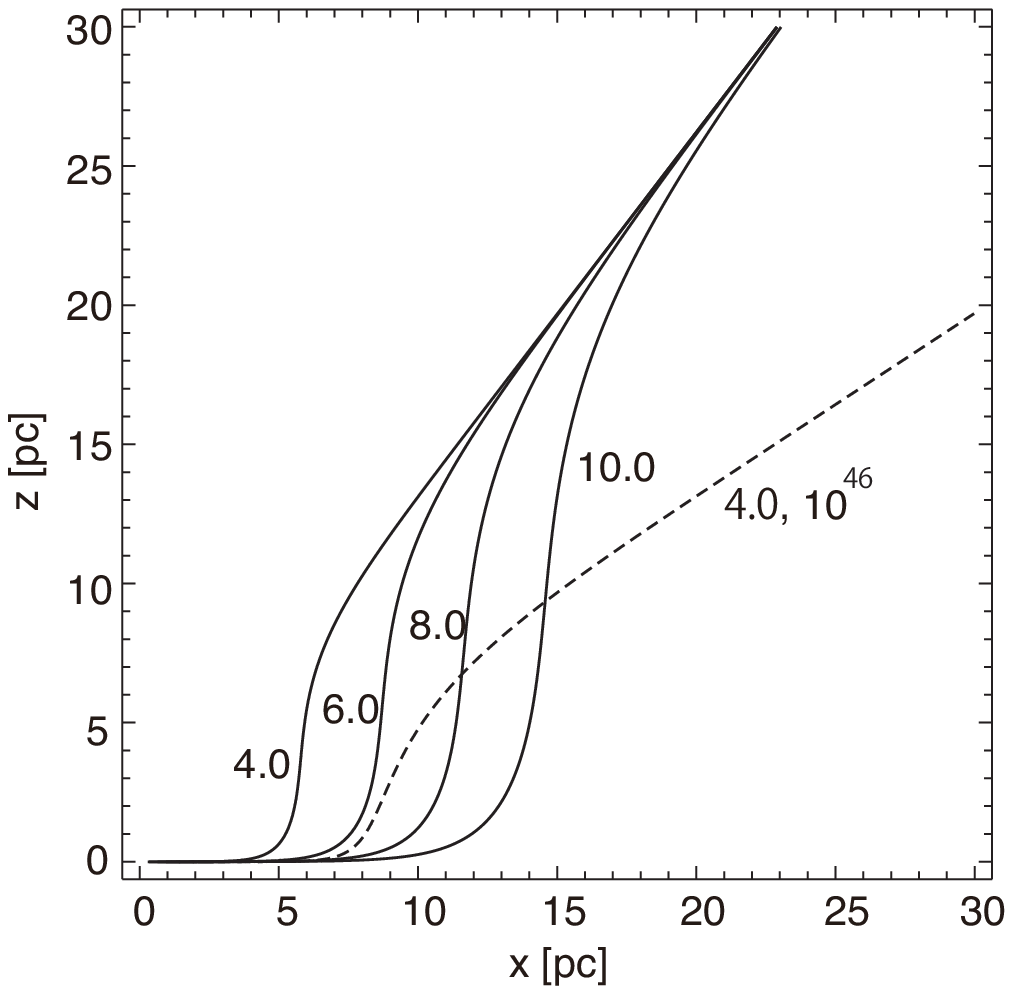} 
\caption{{Critical lines between radiation-pressure dominant and gravity dominant regions for gas disk with different vertical extentions, 
represented by $n(z) = n_{0}/(\sqrt{2\pi} h) e^{-z^{2}/2h^{2}}$ with $h =$ 4, 6, 8, and 10 pc. 
The dashed line indicates the
critical line for $h = 4$ pc and $L_{AGN} = 10^{46}$ erg s$^{-1}$,
which is 100 times that for the other cases.}}

\label{wada_fig: f2b}
\end{figure}

The downward flows can be naturally expected due to the balance between radiation pressure and
the gravity. The ratio between
the radiation pressure for the dusty gas and the gravity due to the central BH is
\begin{eqnarray}
\frac{f_{rad}}{f_{gravi}} = \frac{L_{AGN} \chi_{d} |\cos \theta| }{4\pi G c M_{BH}}e^{-\tau_{d}},
\label{eqn: fr-fg}
\end{eqnarray}
where $\tau_{d}$ denotes the optical depth for the dusty gas at a given point.
Suppose that the gas density is a function of $z$, e.g., $n(z) = N_{col}/\sqrt{2\pi}h e^{-z^{2}/2h^{2}}$.
{ Fig. \ref{wada_fig: f2b} shows
how the radiation-pressure-dominated region and gravity-dominated region are distributed 
around the AGN for the given gas density. 
Here, we assume the dust-to-gas mass ratio of 0.03 \citep{ohs01}, $L_{AGN}=10^{44}\; {\rm erg}\, {\rm s}^{-1}$, dust extinction coefficient $\chi_{d} = 10^{5}\; {\rm cm}^{2} \; {\rm g^{-1}}$,
and the gas density at $z=0$ as $n_{0} = 1000$ cm$^{-3}$.
The four curves represent the condition ${f_{rad}}/{f_{gravi}} = 1$, with the vertical extent of the gas $h = 4,6,8$, and 10 pc, 
for which radiation pressure dominates in the inner side of the curve, therefore outflows are expected.
On the other hand, the radiation pressure cannot prevent gas accretion outside the
critical lines. In fact, accretion flows appear in the
gravity-dominated domain as seen in Figs. \ref{wada_fig: f1} and \ref{wada_fig: f2}.
Moreover, since the critical lines dividing the two domains are not purely radial 
(Fig.  \ref{wada_fig: f2b}), 
outflows with large angles from the $z$-axis may
collide with the boundary of the cavity.}
This pushes the boundaries of the outflow outward, and the gases
 in the outflows change their directions and fall towards the center as indicated by the velocity vectors in Fig, \ref{wada_fig: f2}.
This also causes the biconical cavities to become non-steady, and 
the vertical circulation of the flow, or the  ``fountain'' of gas, circulates
around the AGN on a scale of 10 pc.
The kinetic energy of this circulating flow is partially used to generate 
the random motion in the disk and maintain its thickness.
In other words, both radiation energy and gravity are the cause of origin of the internal turbulent motion
in the disk\footnote{{In a self-gravitating, multi-phase gas disk, gravitational and thermal instabilities themselves can generate 
and maintain turbulent motion \citep{wada2002}. In the present model, 
random motions of the high-density gas near the equatorial plane ($z \sim 0$) 
are dominated by this gravity-driven turbulence. However, the random motions along the vertical direction in
the less dense gas in the thick disk are not driven by gravitational instability.}}.

%
\subsection{Structures and Dynamics in Quasi-steady State}
%

\begin{figure}[t]
\centering
\includegraphics[width = 11cm]{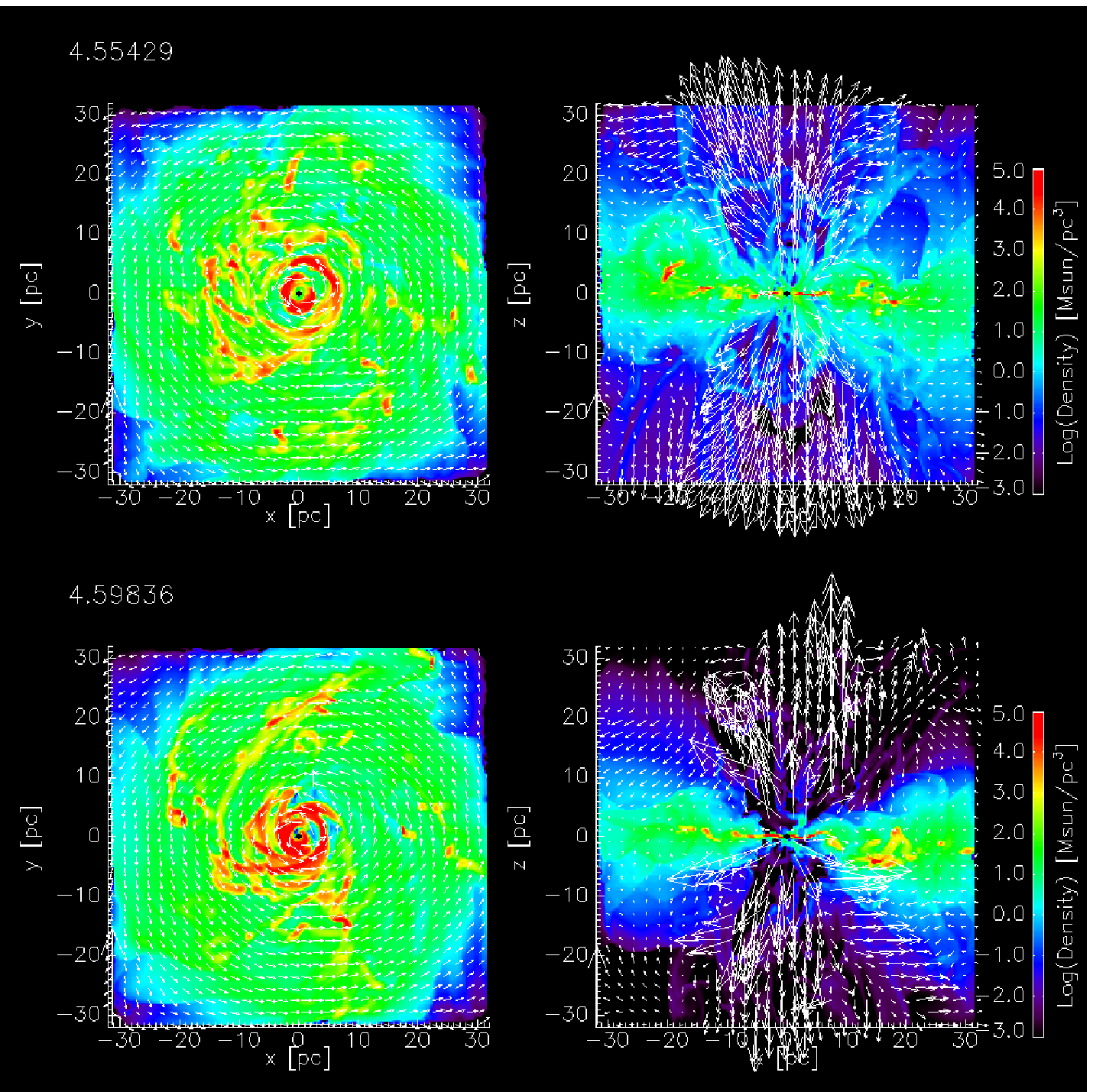}  
\caption{Gas density in quasi-steady state of two models: (top) $L_{AGN}/L_{E} = 0.1$ at $t=4.55$ Myr and (bottom) $L_{AGN}/L_{E} = 0.01$ at $t=4.59$ Myr.
{The vertical slices indicate the $x$-$z$ planes.}}
\label{wada_fig: f4}
\end{figure}

\begin{figure}[t]
\centering
\includegraphics[width = 11cm]{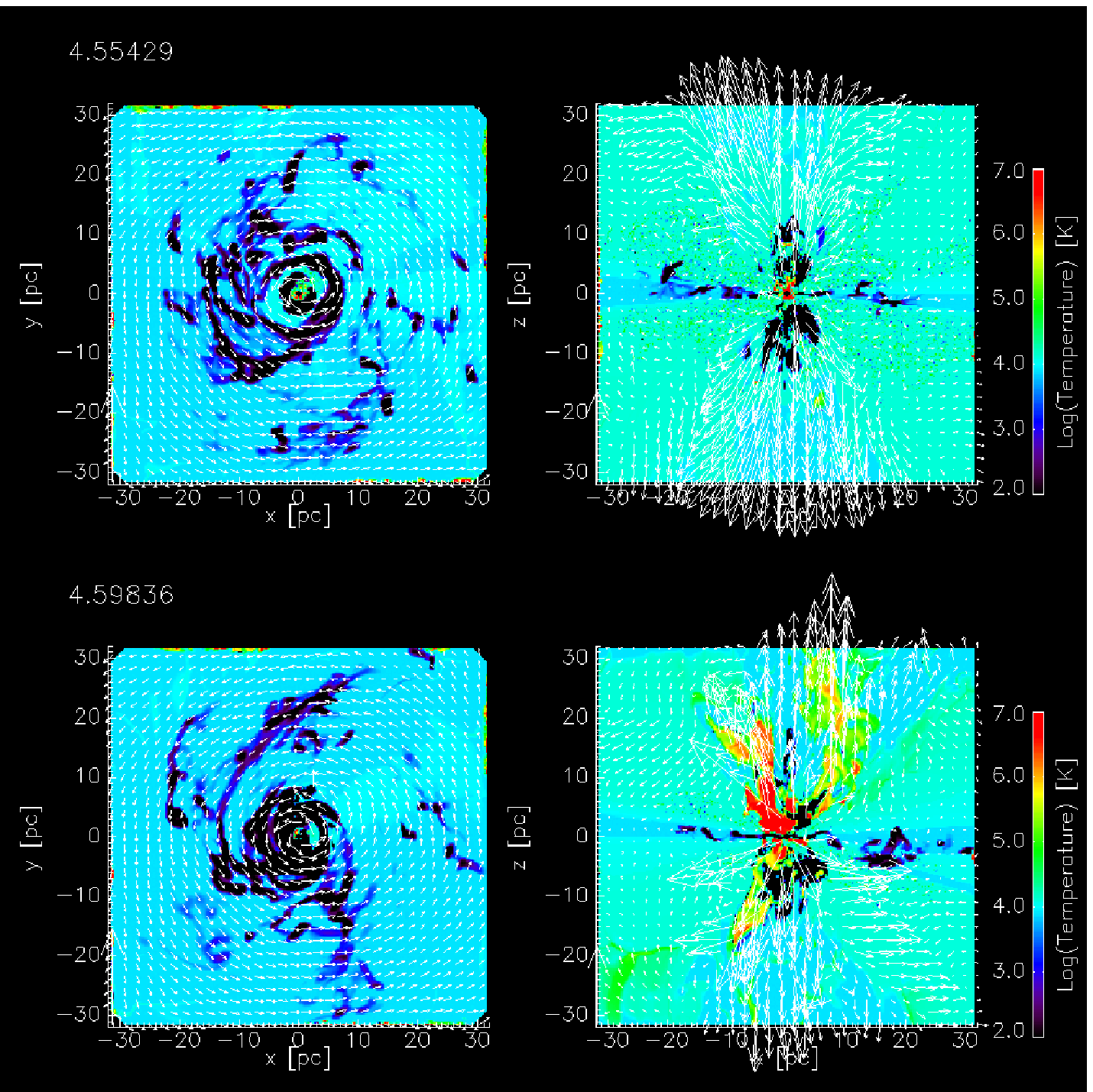}
\caption{Gas temperature in quasi-steady state of two models: (top) $L_{AGN}/L_{E} = 0.1$ at $t=4.55$ Myr and (bottom) $L_{AGN}/L_{E} = 0.01$ at $t=4.59$ Myr.
{The vertical slices indicate the $x$-$z$ planes.}}
\label{wada_fig: f4_tem}
\end{figure}

\begin{figure}[t]
\centering
\includegraphics[width = 7cm]{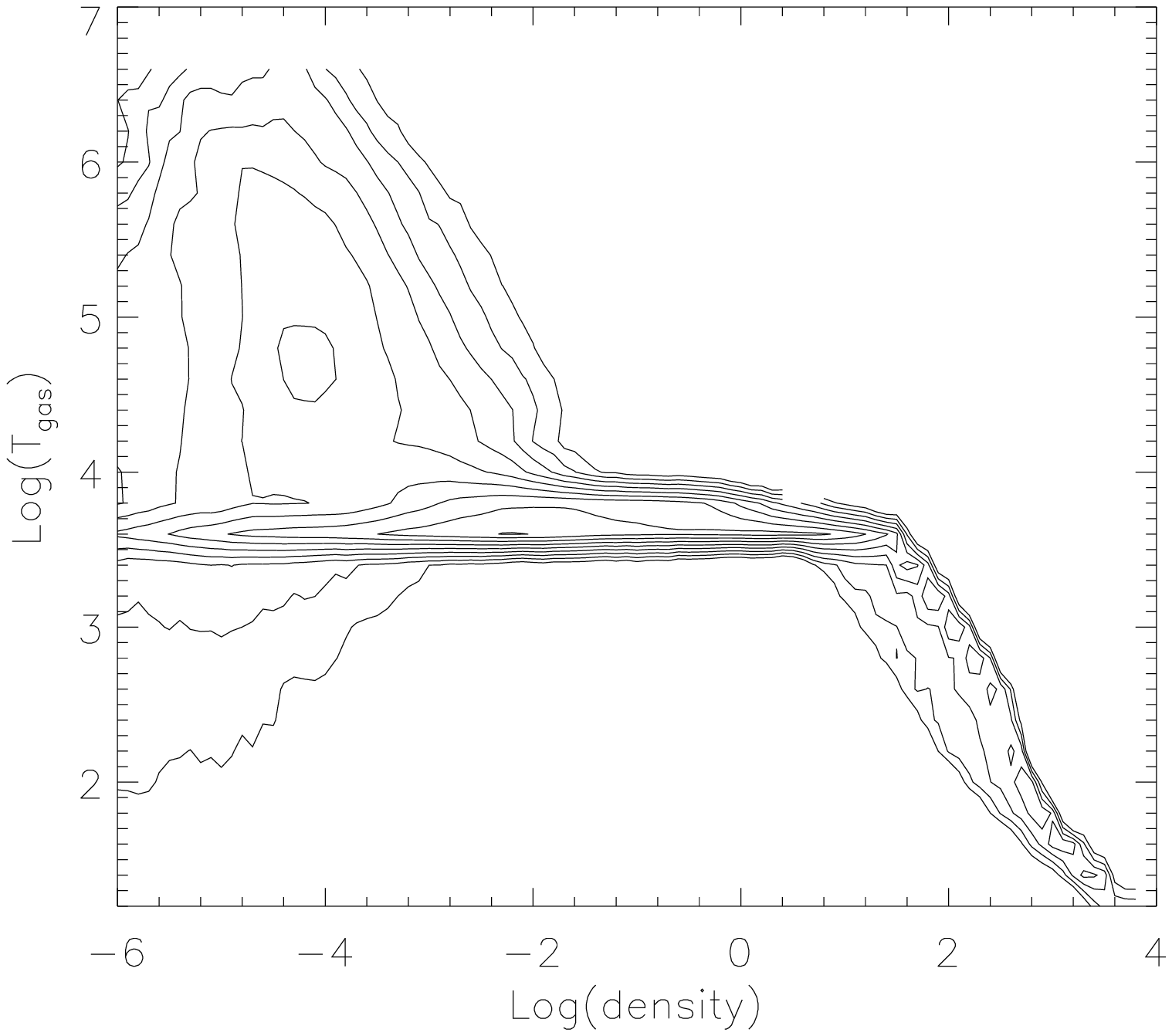} \\
\includegraphics[width = 7cm]{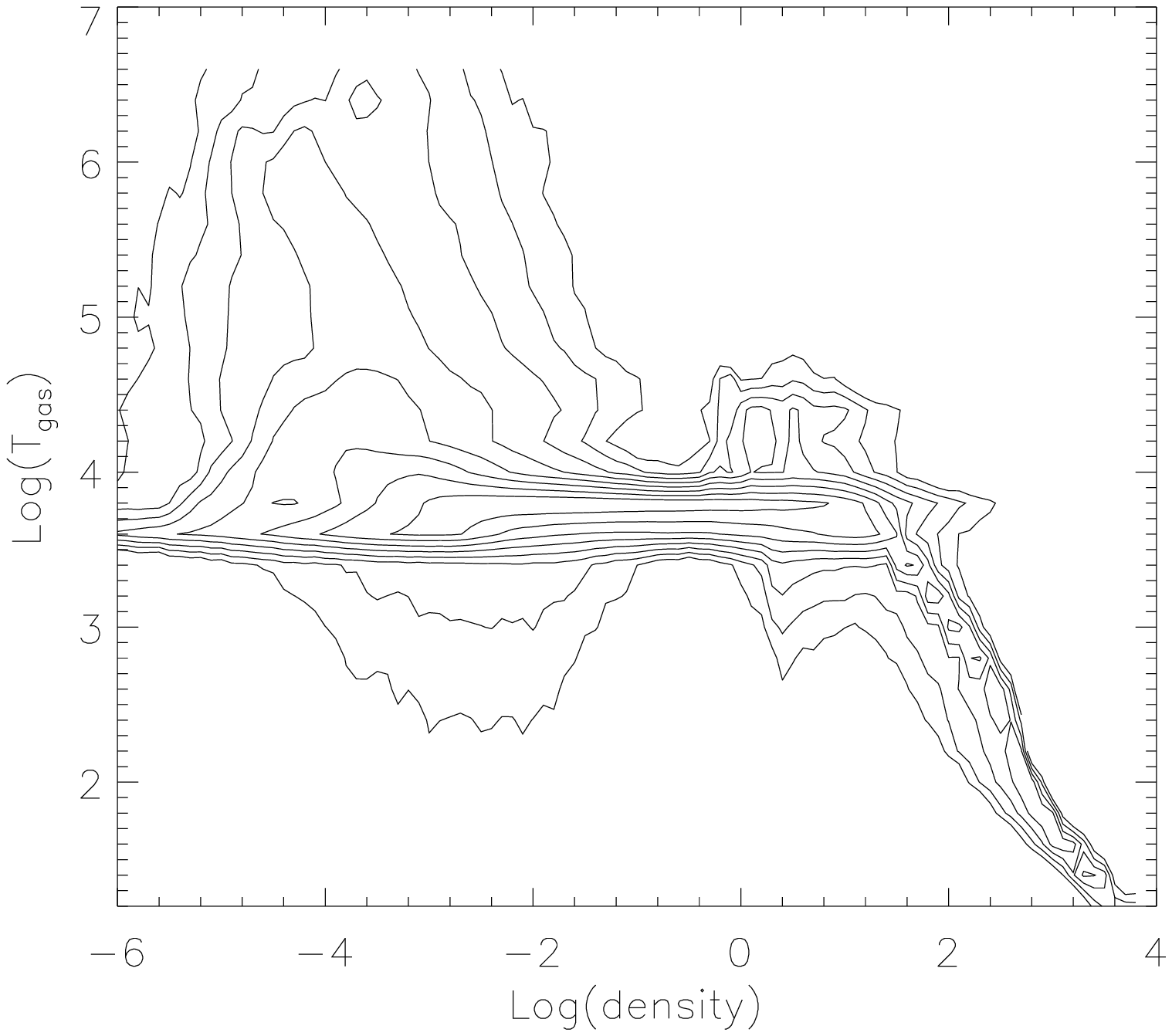} \\
\includegraphics[width = 7cm]{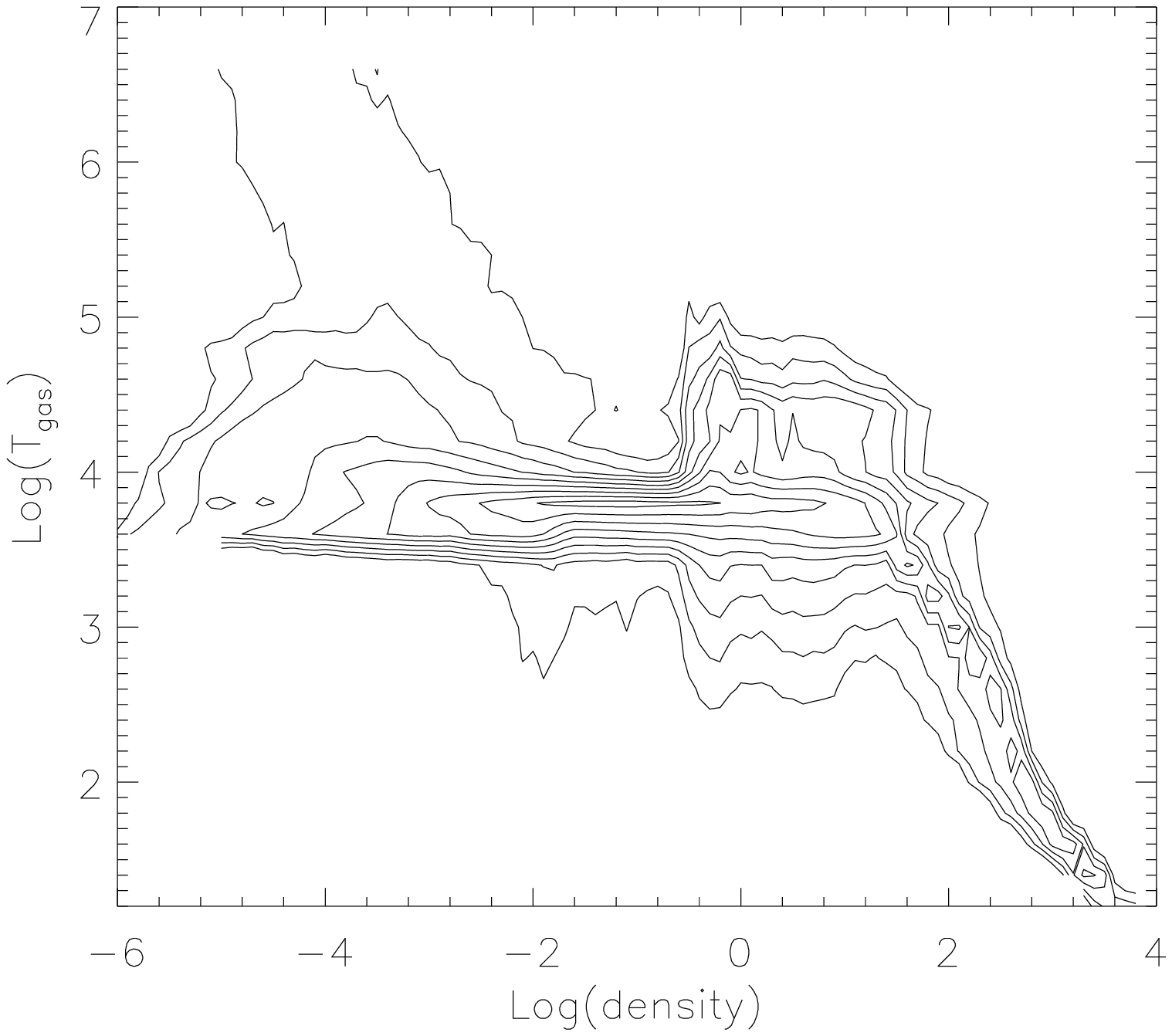} 
\caption{Volumes in gas density [$M_{\odot}\; {\rm pc}^{-3}$] and temperature [K].  (top) no radiation feedback,  (middle) $L_{AGN}/L_{E} = 0.01$, (bottom) $L_{AGN}/L_{E} = 0.1$. }
\label{wada_fig: phase-diagram}
\end{figure}

Figure \ref{wada_fig: f4} shows that the density structures in a quasi-steady states
differ depending on the luminosity of the central source. 
{Here ``quasi-steady'' means that the global morphology has reached an approximately steady state; that is the phase diagram of the gas (Fig. \ref{wada_fig: phase-diagram}) does not 
significantly change after this point. The accretion rates to the central region
are also approximately constant. This is achieved roughly after $t = 3.0-3.5$ Myr
 (Fig. \ref{wada_fig: f6}). Note that since the duration of the present 
 simulation is much shorter than the accretion time scale ($\sim 10^{9}$ Myr), 
 the `quasi-steady' does not necessary mean the long-term stability.
}

In the more luminous model ($L_{AGN}/L_{Edd} = 0.1$, upper panels), inhomogeneous
outflows with $\rho \sim 0.1 - 1\, M_{\odot}\, {\rm pc}^{-3}$ and velocity of $\sim 100$ km s$^{-1}$ are formed, and 
diffuse gas ($\rho \sim 0.1\, M_{\odot}\, {\rm pc}^{-3}$) extends for over 20 pc
above the equatorial plane. 
On the other hand, diffuse and relatively stable 
outflows with $\rho \sim 10^{-2}\, M_{\odot}\,  {\rm pc}^{-3}$ or less
are formed in the less luminous model with $L_{AGN}/L_{Edd} = 0.01$. 
The gas temperature distributions of the two models (Fig. \ref{wada_fig: f4_tem}) show
that dense spiral arms and clumps in the equatorial plane exhibit a temperature of less than 1000 K,
in contrast to the ``warm'' ($\sim 10^{4-5}$ K) torus. 
In the less luminous AGN model, an incresed number of 
hot gas ($\gtrsim 10^{6}$ K) regions appear in the biconical 
outflow. This is because the radiative cooling is less effective 
due to reduced gas density in the region, and as a consequence, the X-ray is 
not attenuated.

Figure \ref{wada_fig: phase-diagram} shows the effects of the radiative feedback on
the phases of the gas. 
The dominant phase is the one around $T_{gas} \sim 8000$ K.
The dense and cold gas ($\rho > 10^{2} \,M_{\odot}\, {\rm pc^{-3}}$
and $T_{gas}  < 1000$ K) is not affected by the radiative feedback.
The X-ray heating produces warm ($T_{gas}  < 1000$ K) gas in
the outer regions (i.e., high latitude) of the thick disk,
as is expected from the XDR model \citep{jp2011}. Shock-heated gas with a temperature of several $10^{4}$ K also 
appears in the same density region.

\begin{figure}[t]
\centering
\includegraphics[width = 11cm]{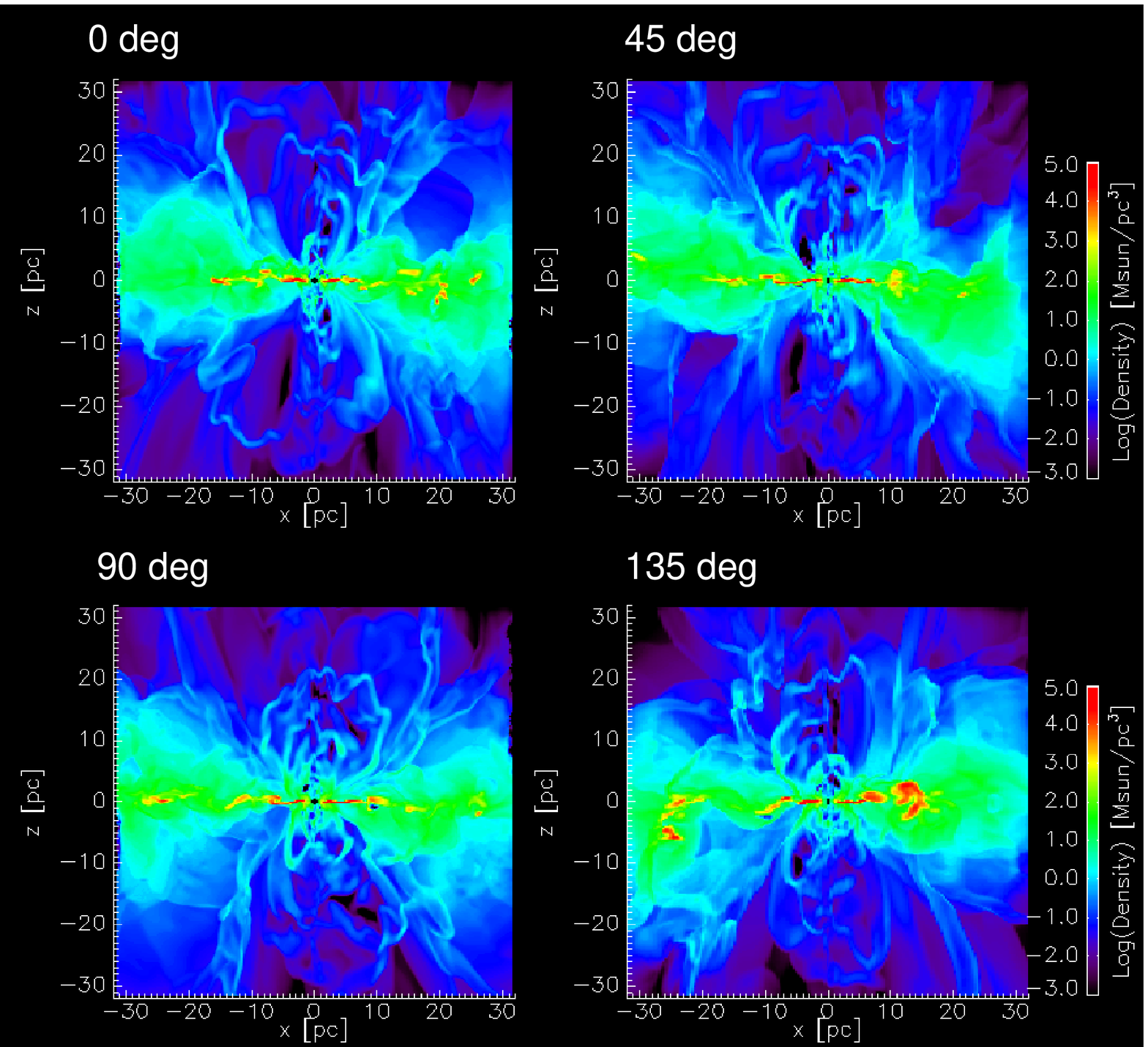}  
\caption{Azimuthal change of gas density distribution of model with $L_{AGN}/L_{E} = 0.1$ at $t= 4.59$ Myr.
In each panel, x-axis is inclined at 0$^{\circ}$, 45$^{\circ}$, 90$^{\circ}$, and 135$^{\circ}$ to the original $x$-axis of the computational box. }
\label{wada_fig: rho-azimuth}
\end{figure}

{ The structure of the torus does not show perfect symmetry for the rotational axis and the equatorial plane.
Figure \ref{wada_fig: rho-azimuth} shows four different vertical slices of density around the rotational axis of 
the model shown in the upper panel of Fig. \ref{wada_fig: f4}.
The diffuse part of torus ($\rho \sim 10^{2}$) is slightly tilted,
thereby indicating that the outlying region of the disk is not completely settled.
The shell-like outflows also show asymmetric distributions.
It is interesting to note that these structures are caused by 
interactions between the radiation from the central symmetric source and the surrounding inhomogeneous gas, thereby
demonstrating importance of the three-dimensional treatment of radiation hydrodynamics in understanding the structure of AGN.
}


\begin{figure}[t]
\centering
\includegraphics[width = 7cm]{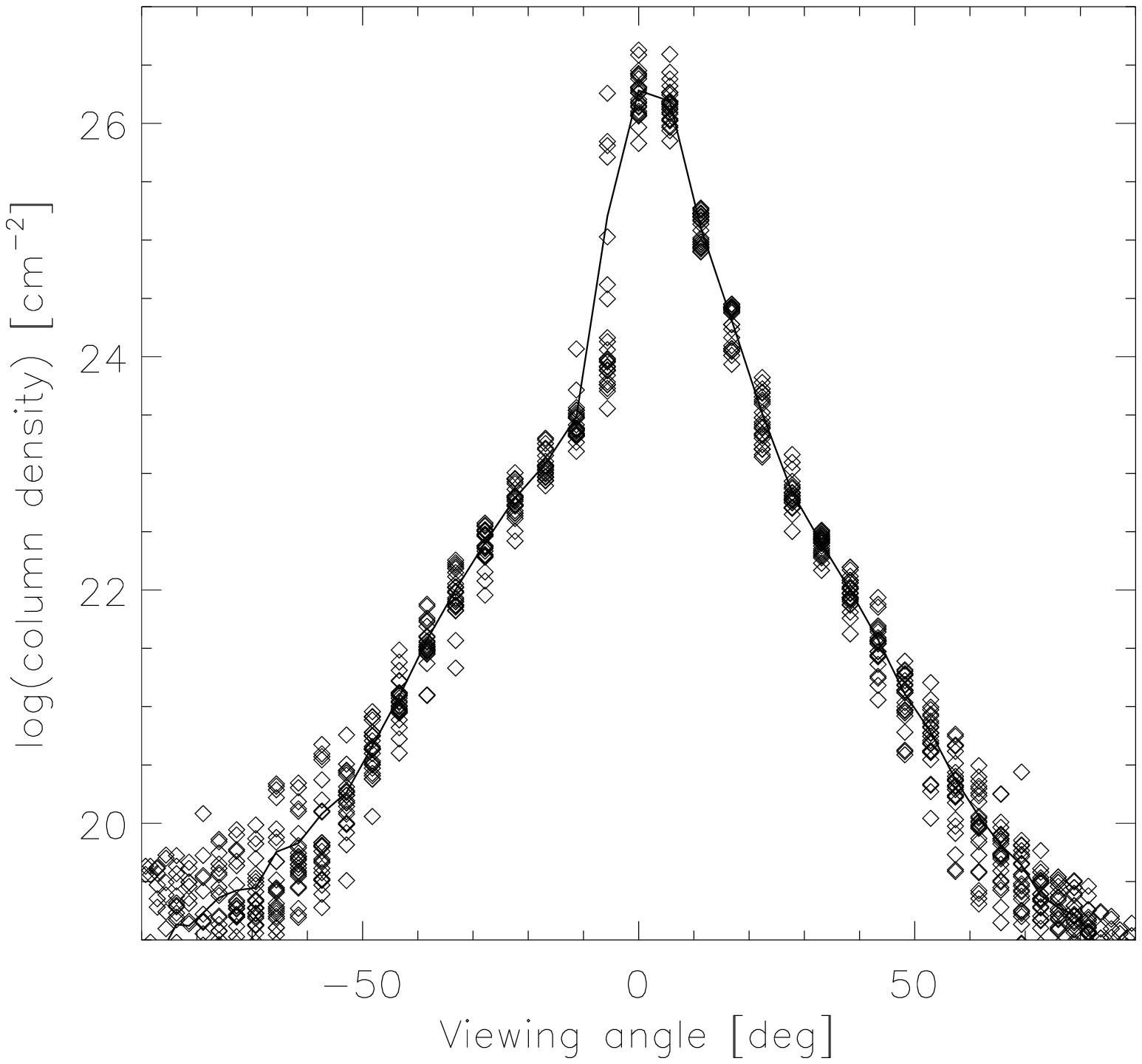} \\ 
\includegraphics[width = 7cm]{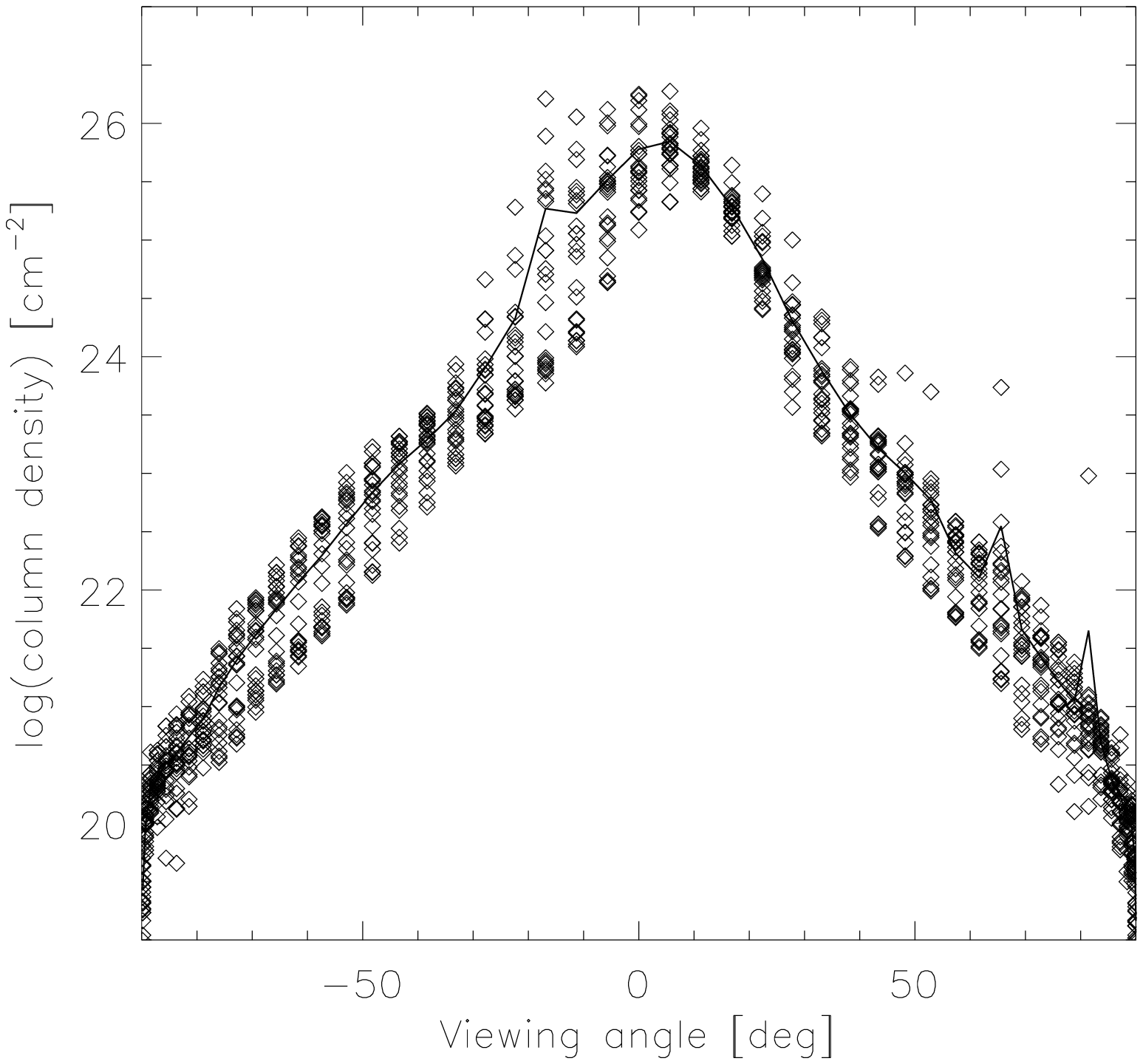} \\ 
 \includegraphics[width = 7cm]{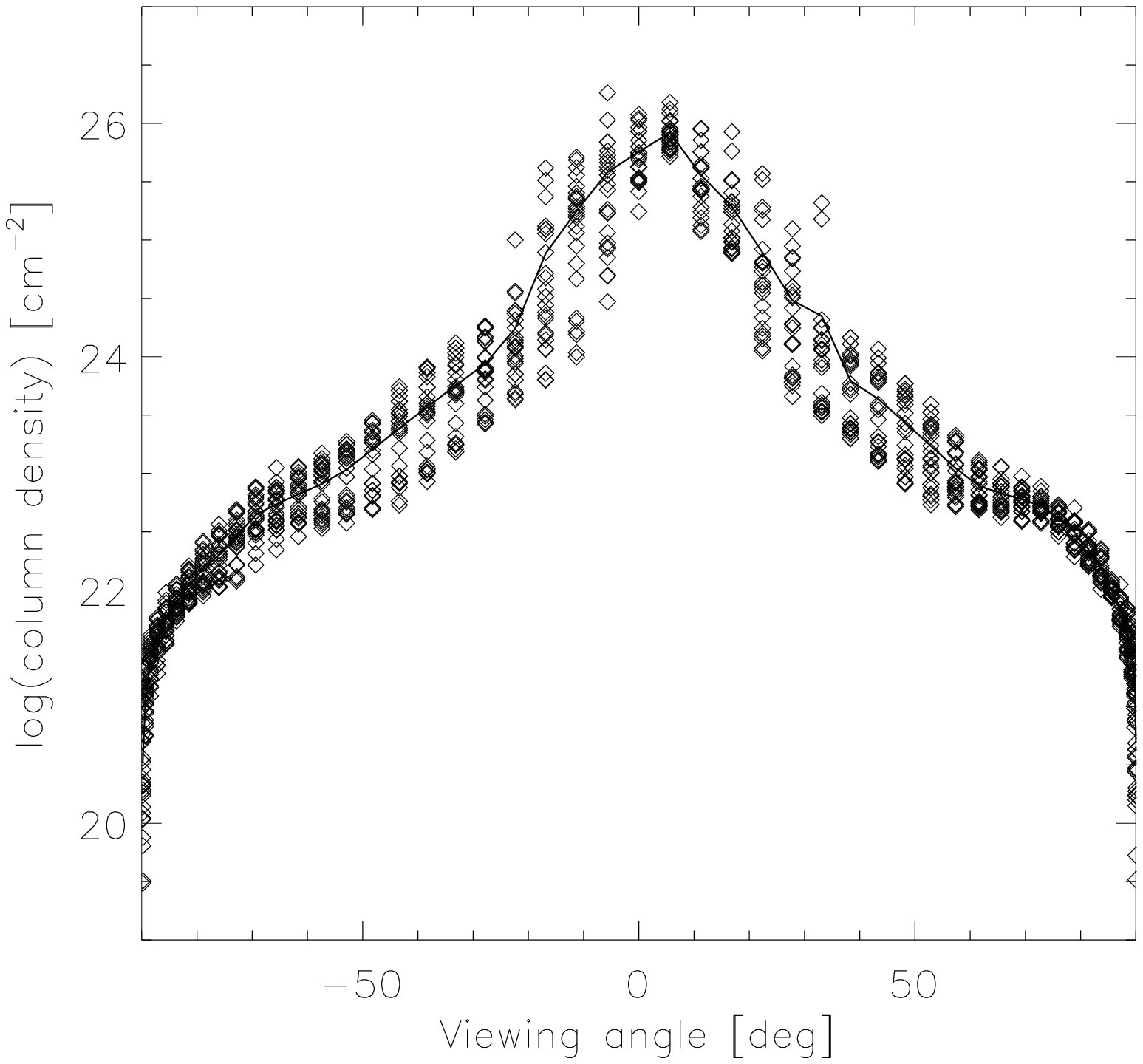} 

\caption{{Total column density distribution as function of viewing angles ($90^{\circ}$: pole-on view and 0$^{\circ}$:
edge-on view) and azimuthal angles.
For each viewing angle, the column density is plotted for 50 azimuthal angles.
The solid line indicates the azimuthal averaged column density for the viewing angles.   (top) Before the radiation feedback
is turned on ($t=1.67$ Myr), (middle) $L_{AGN}/L_{E} = 0.01$ at $t= 4.55$ Myr, (bottom) $L_{AGN}/L_{E} = 0.1$ at $t=4.68$ Myr. }}
\label{wada_fig: f5}
\end{figure}

Obscuration due to the vertically extended gas in the central region over tens of 
parsecs and how it differs between 
models are shown in Fig. \ref{wada_fig: f5}, which shows the plot of the column density toward the central BH as a function of the viewing angle. If there is no radiation from the central source,
the opening angle for $N \leq 10^{23}$ cm$^{-2}$ is approximately 140$^{\circ}$.
The radiation feedback causes the thickness of the disk to increse, and the opening angles are approxiamtely 60$^{\circ}$ and 100$^{\circ}$ 
for $L_{AGN}/L_{E} = 0.1$ and 0.01, respectively.
It is noteworthly that the column density does not only depend on the viewing angle, but also on the
azimuthal angles. The column density varies by a factor of ten
or more for a given viewing angle, thereby indicating the internal inhomogeneous structure of the disk. 
The structure of the disk is similar to that generated by supernova explosions \citep{wada02, wada09}.
However, the present simulations imply that the geometrically and optically thick torus can be naturally formed
due to gravitational energy in the central part with the aid of the radiation feedback
in the case of the AGNs with low Eddington ratios.


The radiation pressure caused by the central source usually has ``negative'' effect on the mass accretion toward the center; however, our results indicate that
mass accretion is not terminated by radiative feedback.
Figure \ref{wada_fig: f6} shows the time evolution of the mass of the two models in four radial bins.
The lower luminosity model ($L_{AGN}/L_{Edd} = 0.01$ ) shows continuous {\it net} mass accretion in the central 8-pc region,
and there is almost no net inflow beyond $r \sim 4 $ pc. 
{The mass accretion in the central part is caused by gravitational torque due to 
the non-axisymmetric features near the equatorial plane of 
the central part of the disk (Fig. \ref{wada_fig: f1})}.
\footnote{
{The amount of the angular momentum 
transfer during one rotational period by $m$-armed spiral density waves is roughly proportional
to $\pi \tan \theta_{i} A_{m}^{2}/m$ \citep[e.g.][]{BT08}, where $A_{m}$ is a ratio of the amplitude of the
spiral to the average density, and $\theta_{i}$ is the pitch angle of the spiral.
For non-linearly developed spiral waves (i.e. $A_{m} \gtrsim 1$) with $\tan \theta_{i} \sim 0.1$, $m \sim 1$, 
about 3\% of the initial angular momentum of the disk is transferred in a rotational period.} }

The accretion rate to the central parsec increases with decreasing AGN luminosity: 
 $\dot{M} = 1.7\times 10^{-3} M_{\odot}\, {\rm yr}^{-1}$ and $\dot{M} = 1.3\times 10^{-4} M_{\odot}\, {\rm yr}^{-1}$ for 
$L_{AGN}/L_{E} = 0.01$ and 0.1, respectively.
For $L_{AGN} = 0.1 L_{E}$,  the net accretion rate is one order of magnitude {\it smaller} than
the accretion rate required to constant AGN luminosity; 
the AGN luminosity is assumed to be constant during the
calculations provided that the energy conversion efficiency is 0.1.
On the other hand, the observed accretion rate in the low luminosity AGN model is comparable
to that required for the luminosity of the source.
These results may suggest that 
luminous AGNs are intermittent, and their luminosity is not instantaneously determined 
by mass inflow from 
$r \sim 1$ pc.

\begin{figure}[t]
\centering
\includegraphics[width = 6cm]{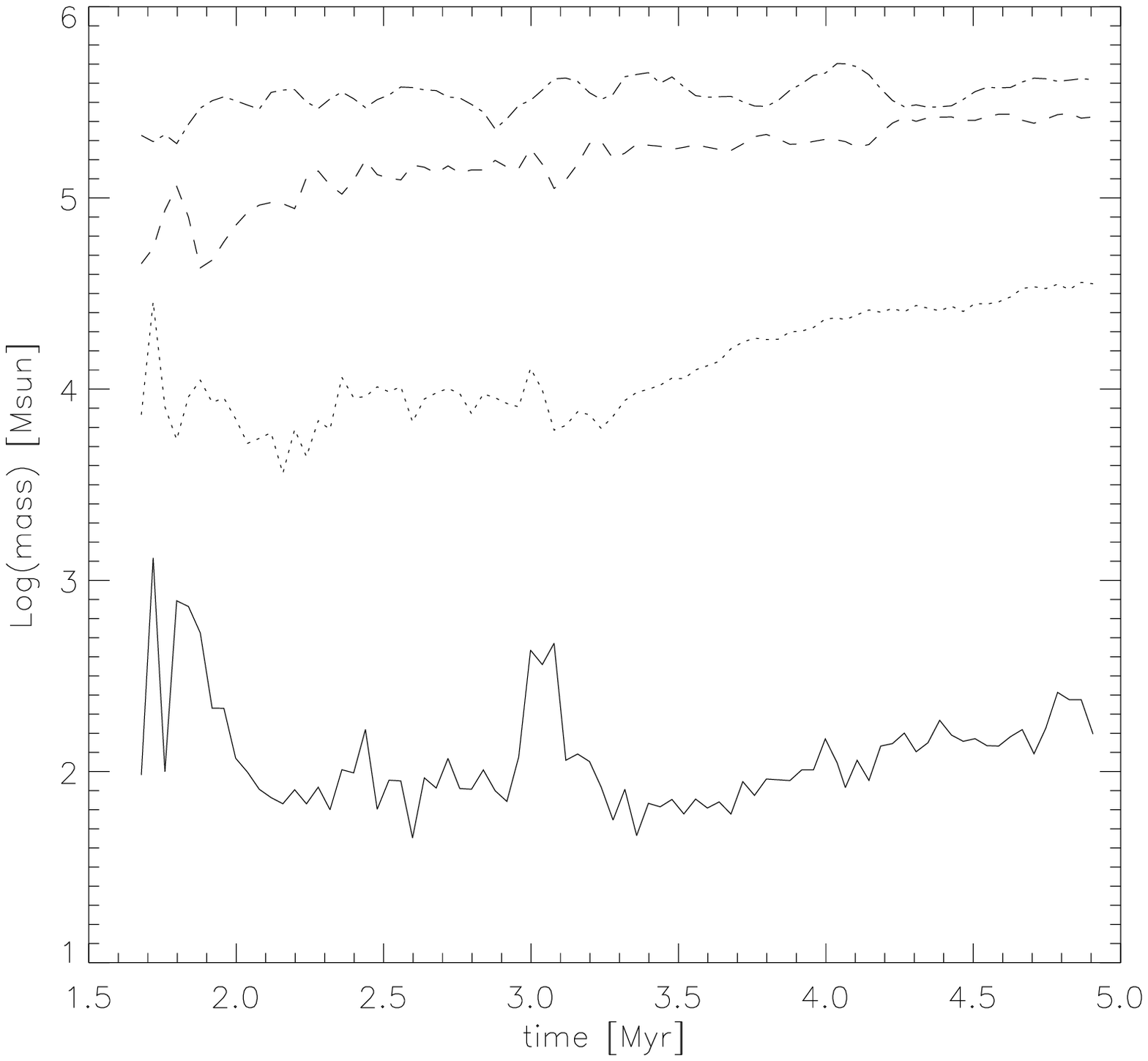} 
\includegraphics[width = 6cm]{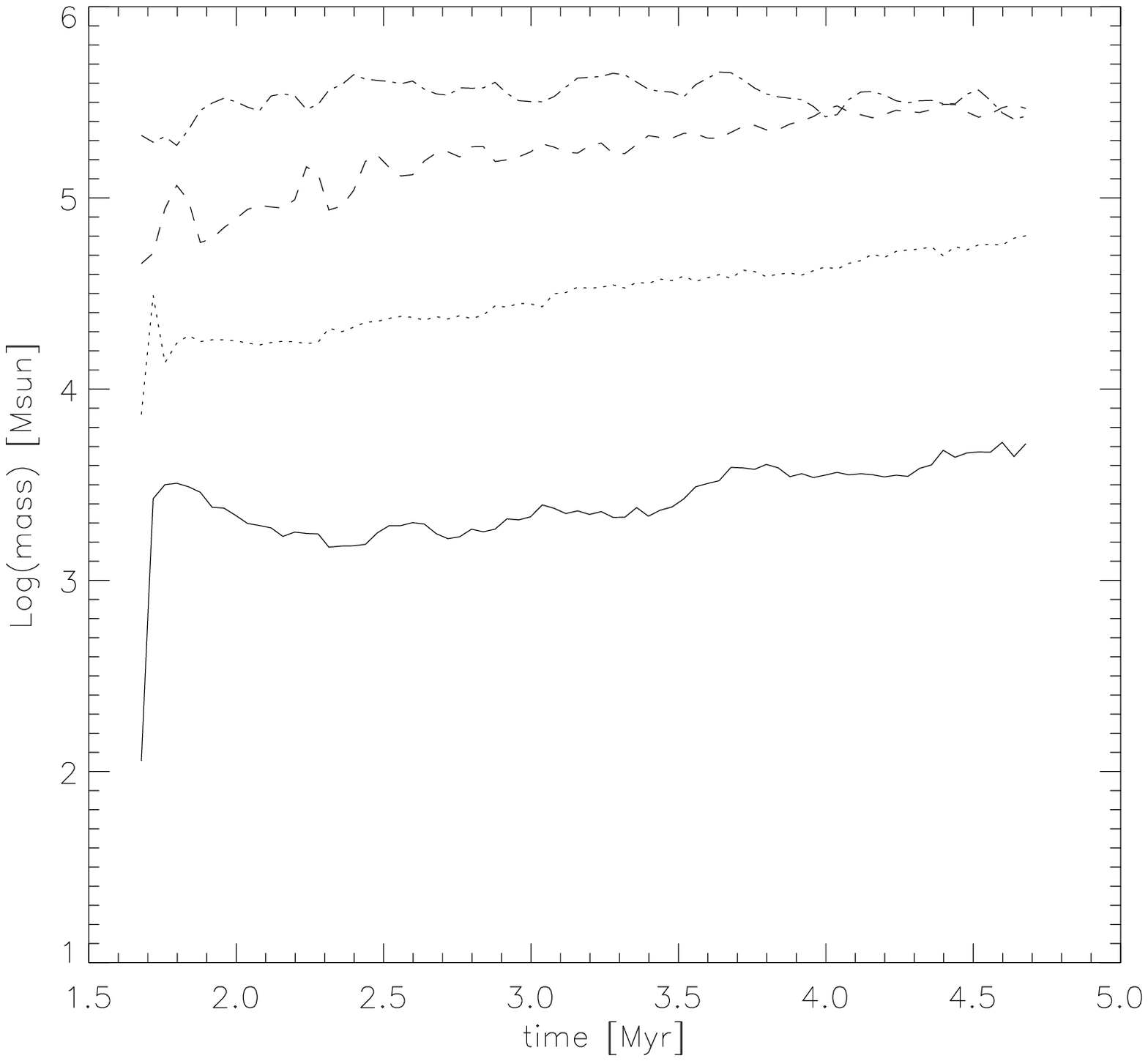} 
\caption{Evolution of radial mass. (left) model qi (Eddington ratio $=0.1$) and (right) model qg (Eddington ratio $=0.01$) for $r<1.0$ pc (thick line), $1\le r < 2$ pc (dotted line), $2\le r < 4$ pc (dashed line), and $4\le r < 8$ pc (dot-dashed line).}
\label{wada_fig: f6}
\end{figure}

%

\section{CONCLUSION AND DISCUSSION}
In our study, we showed that a geometrically and optically thick torus can be naturally formed in the central region
extending tens parsecs around a low luminosity AGN.
We found that radiation drives a ``fountain'' of gas, i.e. vertical circulation, and this also naturally produces a 
thick, turbulent, and inhomogeneous disk that resembles a torus.
The opening angle of the torus for  $N < 10^{23}$ cm$^{-3}$ is
$60-100^{\circ}$ for $L_{AGN} = 0.1-0.01 L_{E}$, in contrast to 140$^{\circ}$ for the model without radiative feedback.
The average accretion rate to the central parsec region is larger in a less luminous AGN; however, this accretion rate is insufficient to 
maintain the AGN luminosity over several Myrs.
These results imply that the AGN luminosity and structure of the surrounding ISM 
alternate between an active phase (high luminosity and a torus with a small opening angle) and inactive phase (low luminosity and a torus with a large opening angle). 
For thick tori (Fig. \ref{wada_fig: f4}) to maintained their structures 
for several million years or longer, 
there should be other mechanisms to enhance the accretion rate, such as supernova-driven turbulence \citep{wada02,wada09}, stellar mass loss \citep{schartmann2010}, turbulence drives by galactic inflows \citep{hopkins2012} and radiation drag \citep{kawakatu2002}.
In fact, recent observations suggest a tight correlation between star formation rate and 
the Eddington ratio of the central BH \citep{chen09}.

Figure \ref{wada_fig: f5} shows that the covering fraction of the central source is 
smaller in AGNs with lower luminosity. This is because more mass can be ejected from the
central region to higher latitudes, and as a result, the radiation from the AGN is
self-obscured. Therefore, the radiation-driven outflows appear only in a small solid angle 
around the rotational axis, as explained in \S 3.1 and illustrated in Fig. \ref{wada_fig: f2b}.
This means that more luminous AGNs are obscured over larger solid angles, 
although this may appear inconsistent with the observed ``receding torus'' \citep{lawrence1991, ueda2003}.
For example, \citet{hasinger2008} studied 1290 AGNs and found that the absorbed fraction {\it decreases} with X-ray luminosity
in the range of $L_{X}=10^{42-46}$ erg s$^{-1}$.
However, it is to be noted that the AGNs explored in the present simulation have
relatively low luminosity, i.e., in the range of $L_{X}=10^{42-43}$ erg s$^{-1}$. 
In this small luminosity range, the structure of the torus does not vary significantly, 
especially for higher values of column density ($N > 10^{24}$ cm$^{-2}$ ). 
For a central source that is 100 times brighter, the radiation-dominant region is 
considerably larger, as shown by the dashed line in Fig. \ref{wada_fig: f2b}, and therefore,
we can expect a larger opening angle for this case.
More recently, \citet{lu2010} have reportred results
that are significantly different from those of 
previous studies; in their SDSS/FIRST survey,  they found that the
type-1 fraction exhibits a constant of ~20\% in the [O III] 5007 luminosity range of $40.7< \log(L_{[{\rm O}_{\rm III}]}/ {\rm erg}\, {\rm s}^{-1}) <43.5$. They also found that the type-1 fraction is independent of the Eddington ratio for its value between 0.01 and 1, if only high density ($> 10 M_{\odot}\; {\rm pc}^{-3}$) gas in the torus contributes to the obscuration; this is also the case as seen 
in Fig. \ref{wada_fig: f4}.

{Further, this study shows that the global structures of the torus and outflows are not static, and 
there is never a perfect symmetry in terms of
 the equatorial plane and the rotation axis  (Fig. \ref{wada_fig: rho-azimuth}),
  even if we assume the presence of angle-dependent, axisymmetric flux from 
the central source. 
The opening angle of the torus and other structures could be 
affected by the choice of the central radiation source. 
In this study, we assumed a thin accretion disk; however,
recent two-dimensional radiation-magnetohydrodynamic simulations 
have shown that geometrically thick and radiatively inefficient accretion flows
associated with outflows are formed depending on the central gas density \citep{ohs11}.
In this case, the radiation flux is collimated around the rotation axis,
and the effect of the radiation on the disk might be less than that observed
in the models used in this study.
Even for a thin accretion disk, the rotational axis is not necessarily aligned to the
galactic rotational axis. This causes anisotropic illumination of the ISM,
and this misalignment could also
affect the occultation of the central source by the torus \citep[see also][]{kawaguchi2011}.
Consequently, we can expect the formation of more such anisotropic structures of the torus and outflows in such cases. This can be confirmed by  
high-resolution observations of nuclear active galaxies associated with 
molecular outflows, such as NGC 1377 \citep{aalto2012b} and by ALMA in the near future.}

%
%
%

%
\vspace{0.5cm}
%
The author is grateful to M. Schartmann for providing 
the dust absorption tables and also for his helpful comments regarding the draft.
We also thank M. Spaans and the anonymous referee for many valuable comments. 
The numerical
computations presented in this paper were carried out on the NEC SX-9 in the
Center for Computational Astrophysics, NAOJ and Cyberscience Center, Tohoku University.
The author thanks N. Horiuchi, Y. Sakaguchi, and the NEC Corporation for 
optimizing the code for the SX-9.
This work is partly supported by Grant-in-Aid for Scientific Research (C) 23540267.




\end{document}